\documentclass[12pt]{article}

\usepackage{latexsym} 
\usepackage{stmaryrd}
\usepackage{epsfig,amssymb,euscript}
\usepackage{amsmath} \usepackage{mathrsfs}
\usepackage{graphicx} \usepackage[latin1]{inputenc}
\usepackage[english]{babel} \usepackage{amsfonts} \usepackage{subfig}
\topmargin=7mm \oddsidemargin=3mm \numberwithin{equation}{section}
\newsavebox{\ns} \newsavebox{\dbrane}
 \def\be{\begin{equation}}
\def\ee{\end{equation}} \def\bea{\begin{eqnarray}}
\def\eea{\end{eqnarray}}

\renewcommand{\theequation}{\arabic{section}.\arabic{equation}}
\def\theequation{\thesection.\arabic{equation}}

 \newcommand\CS{\mathcal{C}}

\newcommand{\cM}{\mathcal{M}}
\newcommand{\cI}{\mathcal{I}}
\newcommand{\cN}{\mathcal{N}}
\newcommand{\bS}{{\bf{S}}}
\newcommand{\bL}{{\bf{L}}}

\newcommand{\bbM}{\mathbb{M}}

\begin{document}

\makeatletter
\renewcommand{\theequation}{\thesection.\arabic{equation}}
\@addtoreset{equation}{section} \makeatother

\baselineskip 18pt

\begin{titlepage}



\vfill

\begin{center}
   \baselineskip=16pt {\Large\bf The Random Discrete Action for
   2-Dimensional Spacetime}
  \vskip 1.5cm {Dionigi M. T. Benincasa${}^a$, Fay Dowker${}^a$,
      Bernhard Schmitzer${}^{a,b}$}\\
   \vskip .6cm
  \begin{small}
      \textit{${}^a$Theoretical Physics Group, Blackett Laboratory, \\
        Imperial College, Prince Consort Rd., London SW7 2AZ, U.K.\\
        ${}^b$Image and Pattern Analysis Group, University of Heidelberg,\\
              Speyerer Str. 6, D-69115 Heidelberg, Germany 
              }
              \end{small}\\*[.6cm]
   \end{center}

\vfill

\begin{center}
\textbf{Abstract}
\end{center}

\begin{quote}
A one-parameter family of random variables, called the Discrete
Action, is defined for a 2-dimensional Lorentzian spacetime of finite
volume. The single parameter is a discreteness scale.  The expectation
value of this Discrete Action is calculated for various regions of 2D
Minkowski spacetime, $\mathbb{M}^2$. When a causally convex region of
$\mathbb{M}^2$ is divided into subregions using null lines
the mean of the Discrete Action is equal to the
alternating sum of the numbers of vertices, edges and faces of the
null tiling, up to corrections that tend to zero as the
discreteness scale is taken to zero.  This result is used to predict that
the mean of the Discrete Action of the flat Lorentzian cylinder is
zero up to corrections, which is verified.  The ``topological''
character of the Discrete Action breaks down for causally convex
regions of the flat trousers spacetime that contain the 
singularity and for non-causally convex rectangles.

\end{quote}

\vfill

\end{titlepage}
\setcounter{equation}{0}

\tableofcontents
\section{The 2-dimensional action of a causal set}

The twin hypotheses that spacetime is fundamentally discrete and that,
of all the structures of classical General Relativity, it is the
causal structure of spacetime that will persist in the deep quantum
regime gives rise to the idea that spacetime is 
a discrete order \cite{Myrheim:1978,tHooft:1979,Bombelli:1987aa}. 
Indeed the basic proposal of the causal set approach to quantum gravity
is that the sum-over-histories for 
quantum gravity is a sum over discrete orders or {\emph{causal sets}}. 
To define such a 
sum-over-histories, it will be necessary to give the amplitude for 
each causal set (or {\emph{pair} of causal sets if 
the path integration is conducted in Schwinger-Kel'dysh manner) and 
 progress has recently
been made on the question of what these
amplitudes might be: a 2-dimensional action and a 4-dimensional action for a
causal set have been proposed \cite{Benincasa:2010ac} and actions in 3, 5 
and higher dimensions
can also be defined \cite{Glaser:2010}. 

Recall that a \emph{causal set} (\emph{causet} for short) is a locally
finite partial order, \emph{i.e.}  it is a pair $(\CS,\preceq)$ where
$\CS$ is a set and $\preceq$ is a relation on $\CS$ which is
reflexive  ($x\preceq x$), acyclic ($y\preceq x\preceq y
\Rightarrow y=x$) and transitive ($z\preceq y\preceq x\Rightarrow
z\preceq x$). Local finiteness is the condition
that the cardinality of any {\emph{order interval}} is finite, where
the (inclusive) order interval between a pair of elements $y\preceq x$
is defined to be $I(x,y) := \{z\in \CS \,|\, y\preceq z \preceq x\}$.
We call $x$ the {\emph{top element}} and $y$ the {\emph{bottom element}}
 of $I(x,y)$.
We write $y \prec x$ when $y \preceq x$ and $y\ne x$.  We define
$n(x,y):= |I(x,y)|$ and call a relation $y\prec x$ a \emph{link} if
$n(x,y) = 2$. A \emph{chain} is a totally ordered subset of $\CS$.

\emph{Sprinkling} is a random process that produces a causet 
which  is a
discretisation of a $d$-dimensional, causal, Lorentzian manifold
$(\mathcal{M},g)$.  It is a Poisson process of selecting points in
$(\mathcal{M}, g)$, independently at random, with density $\rho$
so that the expected number of points sprinkled in a region of
spacetime volume $V$ is $\rho V$.  In quantum gravity we expect that
the density is Planckian so that $\rho = l^{-d}$ where $l$ is of order
the Planck length, but in this paper we treat $\rho$ as a parameter to
be varied. This process generates a causet whose elements are
(identified with) the sprinkled points and whose order is that induced
by the manifold's causal order restricted to the sprinkled points.  If
$(\mathcal{M},g)$ is of finite volume, the causet generated is almost
surely finite and so the process defines a probability distribution
$\mathbb{P}_{\mathcal{M},g, \rho}$ on the set of finite causets
({\textit{aka}} the set of finite partial orders).  Henceforth, for
ease of notation, we will drop the explicit reference to the metric
$g$ and refer, for example, to a spacetime as $\mathcal{M}$ and the
probability distribution above as $\mathbb{P}_{\mathcal{M}, \rho}$.

Sprinkling is not a physical process. It plays a purely
kinematical role and expresses the
discrete-continuum correspondence: a causet $\CS$ is well approximated
by a Lorentzian manifold $\mathcal{M}$ if it could have been
generated, with relatively high probability, by sprinkling into
$\mathcal{M}$. In other words, $\CS$ is well approximated by a
manifold $\mathcal{M}$ if 
there exists an embedding $i:\CS  \hookrightarrow\cM$ such that
(i) $x,y \in \CS$, $y\preceq x$ iff $i(y) \in J^-(i(x))$
 and (ii) the number of elements embedded 
in any sufficiently nice, large region
of volume $V$ is approximately $\rho V$. Strictly, this is a
{\emph{conjecture}}, the ``Hauptvermutung'' of the causal set
approach, but it is supported by much evidence including the result
that a distinguishing Lorentzian geometry
is fully determined by its causal
structure and spacetime
volume measure \cite{Hawking:1976fe,Malament:1977,Levichev:1987}. 

We define the 2D action, $S$, of a finite causal set $\CS$ to be
\cite{Benincasa:2010ac} \be \label{eq:action} S[\CS] = N - 2N_1 + 4N_2 - 2N_3 \ee
where $N$ is the cardinality of $\CS$, and $N_m$ is the number of
inclusive order intervals in $\CS$ of cardinality $m+1$.  $N_1$
therefore is the number of links in $\CS$, $N_2$ is the number of
order intervals that are 3-chains (3 element chains) and $N_4$ is the
number of order intervals that are 4-chains plus the number that are
``diamonds" (with two mutually unrelated elements between the top and
bottom elements).  Note that $N_3$ is not the number of subcausets
that are 3-chains but the number of \emph{order intervals} that are
3-chains.  The form of $S$ as an alternating sum of (weighted) numbers
of things is intriguingly reminiscent of certain topological indices.

 The action (\ref{eq:action}) defines an integer valued random
variable, the {\emph{Discrete Action} $\bS_{\mathcal{M}, \rho}$, for
each finite volume spacetime $\mathcal{M}$ and density $\rho$ 
via the sprinkling process:
$\bS_{\mathcal{M}, \rho}$ takes the value $S[\CS]$ with probability
$\mathbb{P}_{\mathcal{M}, \rho}(\CS)$.  We also define the random
variable $\bS_{\mathcal{M}, N}$ which takes the value $S[\CS]$ with the
probability that causet $\CS$ arises in the process of selecting
exactly $N$ elements uniformly at random -- according to the 
spacetime volume measure 
-- from $\mathcal{M}$.  We then have 
\be \langle\bS_{\mathcal{M},
\rho}\rangle = \sum_{N=0}^\infty \frac{(\rho V)^N}{N!} e^{-\rho V}\langle 
\bS_{\mathcal{M}, N}\rangle \ee 
where $\langle \cdot \rangle$ denotes the expected value, 
$V$ is the spacetime volume of
$\mathcal{M}$, and $\frac{(\rho V)^N}{N!} e^{-\rho V} $ is the
probability that $N$ elements are selected in the Poisson process of
sprinkling into $\cM$ at density $\rho$.
                                                                 
 The Poisson distribution gives for the mean, 
\bea\label{eq:meanS}
\langle\bS_{\mathcal{M}, \rho}\rangle = \rho V -2 \rho^2 \int_{ \mathcal{M}} & &d^dy
\sqrt{-g(y)} \int_{{\mathcal{M}}\cap J^+(y)} d^d x \sqrt{-g(x)} \nonumber\\
{}& &\left(
1 - 2\rho V_{xy} + \frac{1}{2}(\rho V_{xy})^2\right) e^{-\rho V_{xy}}
\eea 
where $V_{xy}$ is the volume of the spacetime causal interval, $[x,y]:=
J^+(y)\cap J^-(x)$, between $x$ and $y$
 and $d$ is the dimension of $\cM$.  
This can be understood thus: $\rho\, d^dx \sqrt{-g(x)}$
is the probability that an element is sprinkled in
an elemental volume at $x$ and similarly for $y$;
$e^{-\rho V_{xy}}$, $\rho V_{xy} e^{-\rho V_{xy}}$ or
$\frac{1}{2}(\rho V_{xy})^2 e^{-\rho V_{xy}}$ is the probability
that there is no element, one element or two elements, respectively,
sprinkled in $[x,y]$.

Note that the double integration may be done in either order:
\bea\label{eq:meanSreverse}
\langle\bS_{\mathcal{M}, \rho}\rangle = \rho V -2 \rho^2 \int_{ \mathcal{M}} & & d^dx
\sqrt{-g(x)} \int_{{\mathcal{M}}\cap J^-(x)} d^d y \sqrt{-g(y)} \nonumber\\
{}& &\left(
1 - 2\rho V_{xy} + \frac{1}{2}(\rho V_{xy})^2\right) e^{-\rho V_{xy}}\,.
\eea
Indeed, the causet action 
(\ref{eq:action}) is invariant under
reversal of the order relation on $\CS$, and so 
the Discrete Action (DA) for any spacetime $(\cM, g)$ is 
equal to the DA of its time-orientation-reverse. 

$\bS_{\mathcal{M}, \rho}$ is defined for any finite volume (causal)
spacetime of any dimension so we can ask in what sense it is 2-dimensional.
 Each
realisation of $\bS_{\mathcal{M}, \rho}$ is the action $S[\CS]$ of some
finite causet $\CS$ and 
\be S[\CS] = \sum_{e_i \in \CS}
{L}(e_i) \ee 
where 
\be {L}(e_i) = 1 - 2n_1(e_i) +4
n_2(e_i) - 2n_3(e_i) \ee 
and $n_m(e_i)$ is the number of inclusive
order intervals in $\CS$ with cardinality $m+1$ and with top element
$e_i$. $L(\cdot)$ itself defines a random variable, $\bL_{\mathcal{M},
\rho, y}$, for each spacetime $\mathcal{M}$, each point $y
\in\mathcal{M}$ and each $\rho$ in the following way.  Fix $y\in
\mathcal{M}$, sprinkle into $\mathcal{M}$ at density $\rho$ and add an
element at $y$ to the sprinkled causet to form causet $\CS'$ which has
a marked element, call it $e_y$. The value of $\bL_{\mathcal{M},
\rho,y}$ is then $L(e_y)$ evaluated in $\CS'$.  If $\mathcal{M}$ is 2-dimensional, 
the mean of $\bL_{\mathcal{M}, \rho,y}$ tends to $\frac{1}{4\rho}R(y)$, where $R(y)$ is
the Ricci scalar, as $\rho$ tends to infinity \cite{Benincasa:2010ac}.  It approaches
its limit when the discreteness length scale $l:= \rho^{-\frac{1}{2}}$
is much smaller than the curvature scale $R^{-\frac{1}{2}}$.  If
$\mathcal{M}$ is {\emph{not}} 2-dimensional, there is no apparent
reason for $\bL_{\mathcal{M}, \rho,y}$ to have anything to do with the
continuum geometry $\mathcal{M}$.

Since $L$ is thus related to the Ricci scalar when the causal
set is a 2D sprinkling and $S$ is a sum of $L(\cdot)$ over the
causal set, this implies that when $\mathcal{M}$ is 2-dimensional and
as $\rho \rightarrow \infty$, $\langle \bS_{\mathcal{M}, \rho}\rangle $ will tend to
something that contains a term $\frac{1}{4}\int_{\mathcal{M}} d^2x
\sqrt{-g} R$ plus terms arising from boundary effects.  We will
investigate this and in particular the nature of the boundary terms.
In doing so we will be exploring whether the 2D Discrete 
Action is topological in character. The standard gravitational action for 2D Euclidean
gravity, with its Einstein-Hilbert term and the (2D analogue of the) Gibbons-Hawking
boundary term, is known to be a topological invariant, due to the Gauss-Bonnet theorem.
The Gauss-Bonnet Theorem has been extended to Lorentzian
manifolds \cite{Chern:1963, Avez:1963}, so for ordinary (Lorentzian) 2D gravity, the action with 
an appropriate boundary
term is also topological and a question arises: to what extent is the 2D causal set action
topological?            

\section{Intervals in $\mathbb{M}^2$}

Consider a causal interval in
2D Minkowski spacetime,  $\mathcal{I}:= [p,q] \subset \mathbb{M}^2$.
For definiteness consider the interval to have fixed volume (area), $V$.

Following a conjecture of R. Sorkin, G. Brightwell proved that the
 mean $\langle\bS_{\mathcal{I}, N}\rangle =1$, for any $N\ne 0$ \cite{Brightwell:2009}.  This implies
 that the mean of $\bS_{\mathcal{I}, \rho}$ is 
\bea \langle\bS_{\mathcal{I},
 \rho}\rangle & = \sum_{N=1}^\infty \frac{(\rho V)^N}{N!} e^{-\rho V}\nonumber\\ & =
 1 - e^{-\rho V}\label{eqn:action_minkowski2} \eea 
where $ \frac{(\rho V)^N}{N!}
 e^{-\rho V}$ is the probability, in the Poisson process, that $N$
 elements are sprinkled into $\mathcal{I}$.

We use (\ref{eq:meanS}) to prove this result in a different way:
\be\label{eq:flatS} \langle\bS_{\mathcal{I}}\rangle = \rho V -2 \rho^2 \int_{
\mathcal{I}} d^2y \int_{{\mathcal{I}}\cap J^+(y)} d^2 x\,\,\, p(\rho
V_{xy}) \ee 
where $p(\xi) = (1 - 2\xi +
\frac{1}{2}\xi^2) \exp({-\xi})$ 
and we have suppressed the subscript $\rho$ on the
random variable $\bS_{\mathcal{I}, \rho}$.

We use coordinates in which $p$ and $q$ lie on the time axis and $q$
is at the origin. We consider null coordinates 
$u_x = \frac{1}{\sqrt{2}}(x^0 - x^1)$, 
$v_x = \frac{1}{\sqrt{2}}(x^0 + x^1)$
and similarly for $u_y$, $v_y$. Then the interval is defined by $u, v
\in [0,a]$ for $a = \sqrt{V}$.

\begin{equation}
\langle \bS_{\mathcal{I}} \rangle  =  \rho V -2 \int_0^a du_x \int_0^a
dv_x \int_{0}^{u_x} du_y \int_{0}^{v_x} dv_y\,\rho^2\,p(\rho\,\Delta
u\,\Delta v) \label{eqn:action_minkowski} 
\end{equation}
where $\Delta u = u_x-u_y,\ \Delta
v = v_x-v_y$.
\begin{eqnarray}
\langle \bS_{\mathcal{I}}\rangle
&=&  \rho V -2 \int_0^a du_x \int_0^a dv_x
\int_{0}^{u_x} d\Delta u \int_{0}^{v_x} d\Delta
v\,\rho^2\,p(\rho\,\Delta u\,\Delta v) \nonumber \\ & = & \rho V -2
\int_0^a du_x \int_0^a dv_x \left[ \left[ \text{integrand
1}\right]^{\Delta v=v_x}_{\Delta v=0} \right]^{\Delta u=u_x}_{\Delta
u=0} \nonumber
\end{eqnarray}
where
\begin{eqnarray} 
{} & &{\textrm {integrand 1}} =
-\frac{\rho}{2}\left(1-\rho\,\Delta u\,\Delta
v\right)\,\exp(-\rho\,\Delta u\,\Delta v) \nonumber \\ & &
[g(\xi)]^{\xi=\alpha}_{\xi=\beta}=g(\alpha)-g(\beta). \nonumber 
\end{eqnarray}
Hence
\be
 \langle S_{\mathcal{I}} \rangle  = 
1-\exp(-\rho\,a^2) = 1 - \exp(-\rho V)\,.
\ee
As $\rho \rightarrow \infty$, $\langle S_{\mathcal{I}} \rangle
\rightarrow 1$ and we write $\langle S_{\mathcal{I}} \rangle \approx
1$ to denote this.

Consider now splitting up the interval $\mathcal{I}$ into four smaller
intervals ${\mathcal{I}}_i$, $i = 1,\dots 4$, as shown in 
Fig. \ref{fig:action_minkowski_split}. When computing the expected
value of $\bS_{\mathcal{I}}$ one can split the integral up into the
means of the actions of the four subintervals plus the ``bilocal''
contributions when $x$ and $y$ lie in two different subintervals.
More concretely, given any subcausets, $A$ and $B$ of a causal set
$\CS$, we define the bilocal action, 
\be S[\CS;A,B] = N(A,B) - 2
N_1(A,B) + 4 N_2(A,B) - 2N_3(A,B)
 \ee 
where $N(A,B)$ is the number of
elements in $A\cap B$ and $N_m(A,B)$ is the number of inclusive order
intervals in $\CS$ of cardinality $m+1$ with top element in $A$ and
bottom element in $B$. Now let $X$ and $Y$ be submanifolds of
spacetime $\mathcal{M}$.  We define the random variable,
$\bS_{\mathcal{M};X,Y}$, the {\emph {Discrete Bilocal Action}},
via the sprinkling process:
sprinkle into $\mathcal{M}$ at density $\rho$ \footnote{To simplify
notation, we don't make the dependence on the density explicit.} to
obtain causet $\CS$ with subcauset $A$($B$) being that sprinkled into
$X$($Y$).  For that realisation, $\bS_{\mathcal{M}; X, Y}$ takes the
value $S[\CS;A,B]$.

Note that $\bS_{\mathcal{M}; X, X} = \bS_{X}$ if $X$ is a {\emph{causally
convex}} subset of $\mathcal{M}$.\footnote{A causally convex region, $X$, of
$\mathcal{M}$ is one such that $x,y\in X$ implies that the causal
interval in $\cM$ between $x$ and $y$ is a subset of $X$.}
 
Now, consider $\mathcal{I}$ and its subintervals.  If we adopt
$\bS_{ij}$ as simplified notation for the bilocal action
$\bS_{\mathcal{I}; \mathcal{I}_i, \mathcal{I}_j}$, then we have
\begin{equation}
\label{eqn:action_minkowski_split}
\langle \bS_{\mathcal{I}} \rangle = \sum_{i=1}^4\langle
\bS_{\mathcal{I}_i} \rangle + \sum_{\substack{i,j =1\\ j<i}}^4 \langle
\bS_{ij}\rangle\,.
\end{equation}
\begin{figure}[htb]
	\centering \subfloat[$a$ and $c$ ($b$ and $d$) are the
	$v$-coordinate ($u$-coordinate) lengths of the sides
	of the subintervals]{
		\includegraphics[scale=1.2]{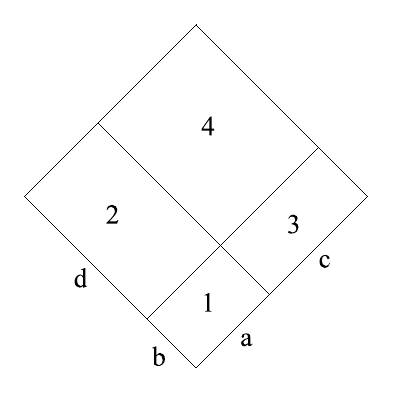}
			\label{fig:action_minkowski_split}
		}
		\hskip 1.5cm \subfloat[]{
		\includegraphics[scale=1.2]{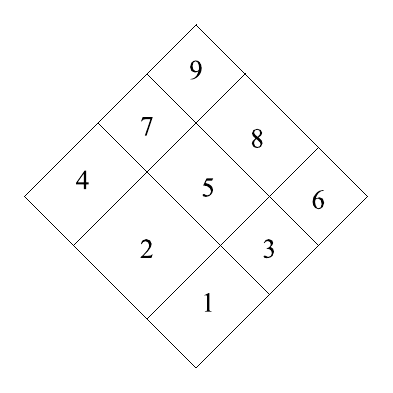}
			\label{fig:action_minkowski_split_b}
		}
	\caption{Splitting up a causal interval in 2D Minkowski to
	compute the action}
	\label{fig:action_minkowski}
\end{figure}
The bilocal summands can be computed using the integral in
Eq. (\ref{eqn:action_minkowski}) and adjusting the boundaries. 
This yields
\begin{eqnarray}
\label{eqn:action_minkowski_21}
\langle \bS_{21} \rangle& = & -2\int_0^a dv_x
\int_b^{b+d} du_x \int_{0}^{v_x} dv_y \int_{0}^{b}
du_y\,\rho^2\,p(\rho\,\Delta u\,\Delta v) \nonumber \\ & = & -2\int_0^a
dv_x \int_b^{b+d} du_x \left[ \left[ \text{integrand 1}
\right]^{\Delta u=u_x}_{\Delta u=u_x-b} \right]^{\Delta v=v_x}_{\Delta
v=0} \nonumber \\ & = &
-1+\exp(-a\,b\,\rho)+\exp(-a\,d\,\rho)-\exp(-a\,(b+d)\,\rho)\\
 &\approx& -1 \nonumber
\end{eqnarray}
and
\begin{eqnarray}
\label{eqn:action_minkowski_41}
\langle \bS_{41} \rangle & = &-2 \int_a^{a+c} dv_x
\int_b^{b+d} du_x \int_{0}^{a} dv_y \int_{0}^{b}
du_y\,\rho^2\,p(\rho\,\Delta u\,\Delta v) \nonumber \\ & = &
-2\int_a^{a+c} dv_x \int_b^{b+d} du_x \left[ \left[ \text{integrand 1}
\right]^{\Delta u=u_x}_{\Delta u=u_x-b} \right]^{\Delta v=v_x}_{\Delta
v=v_x-a} \nonumber \\ & = &
1-\exp(-(a+c)\,(b+d)\,\rho)\nonumber \\ & &\
+\exp(-a\,(b+d)\,\rho)+\exp(-c\,(b+d)\,\rho))\nonumber \\ & &\
+\exp(-(a+c)\,b\,\rho)+\exp(-(a+c)\,d\,\rho)
\nonumber \\ & &\
 -\exp(-a\,b\,\rho)-\exp(-a\,d\,\rho)-\exp(-c\,b\,\rho)-\exp(-c\,d\,\rho)\\
  &\approx& 1\,. \nonumber
\end{eqnarray}
The three other bilocal contributions $\langle \bS_{ij}\rangle$  can be
obtained from $\langle \bS_{21} \rangle$ by changing the
parameters appropriately. Putting together all parts of
Eq. (\ref{eqn:action_minkowski_split}) one exactly recovers
Eq. (\ref{eqn:action_minkowski2}).

Now, one can continue this game and split up the interval
even further as in Fig. \ref{fig:action_minkowski_split_b}. To
compute the mean of the action one must again
calculate 
\begin{equation}
\label{eqn:action_minkowski_split9}
\langle \bS_{\mathcal{I}} \rangle = \sum_{i=1}^9\langle
\bS_{\mathcal{I}_i} \rangle + \sum_{\substack{i,j =1\\ j<i}}^9 \langle
\bS_{ij}\rangle\,.
\end{equation}
We already know the contributions $\langle \bS_{I_i} \rangle \approx
1$ and the bilocal contributions 
from two intervals that either share an edge 
or lie above and below a shared vertex
({\textit{e.g.}} $\langle \bS_{21} \rangle$ and
$\langle \bS_{51} \rangle$ in
Fig. \ref{fig:action_minkowski_split_b}). It remains to compute 
the bilocal contributions from pairs of intervals such as 
(4,1),(7,1) and (9,1) in
Fig. \ref{fig:action_minkowski_split_b}. It turns out they 
consist only of exponential terms that 
are small when intervening intervals are large on the discreteness scale.  
In the limit of large density, we are left 
with a contribution of  $1$ for every
subinterval, $-1$ for every edge and
$1$ for every vertex. One could
write
\begin{equation}
\label{eqn:action_euler}
\langle \bS \rangle \approx F-E+ V
\end{equation}
where $F$ denotes the number of faces
{\textit{i.e.}} intervals, $E$ the number of edges and
$V$ the number of vertices. $F - E+V$ is 
the formula for the Euler character of a 
polyhedron and motivates the question: Is the expected
action (to some extent) a topological invariant? It is obvious that the
formula can be applied to arbitrary causally convex
regions of $\bbM^2$ that can be tiled by
causal intervals as long as each interval is large enough
for the corrections to be negligible.
It is not hard to verify that any
such region will have a mean Discrete Action $\langle \bS\rangle \approx 1$.
So for example the region shown in
Fig. \ref{fig:action_minkowski_construction_1} will give $\langle \bS
\rangle \approx 1$ but the region in
Fig. \ref{fig:action_minkowski_construction_2} will not.
\begin{figure}[htb]
	\centering \subfloat[causally convex]{
		\includegraphics[scale=1.2]{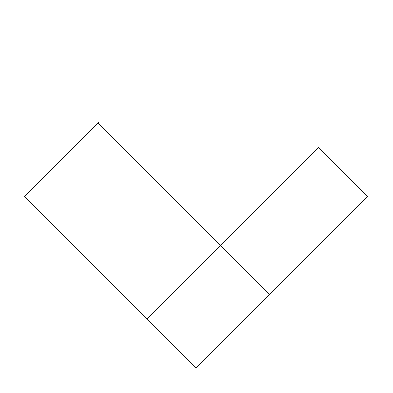}
			\label{fig:action_minkowski_construction_1}
		}
		\hskip 1.5cm \subfloat[not causally convex]{
		\includegraphics[scale=1.2]{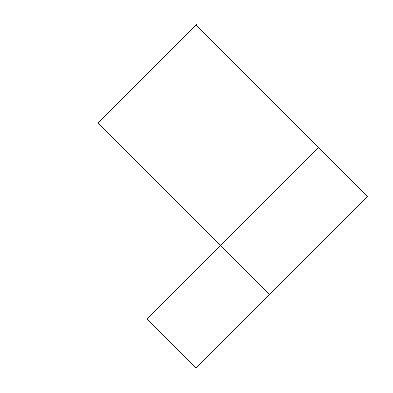}
			\label{fig:action_minkowski_construction_2}
		}
	\caption{Different regions constructed from causal intervals
	in $\mathbb{M}^2$}
	\label{fig:action_minkowski_construction}
\end{figure}

\section{Causally convex regions in $\mathbb{M}^2$}

The boundary of a causally convex region of $\mathbb{M}^2$  can be 
spacelike in parts, but never 
timelike. 
If the region's boundary comprises straight line segments,  
such as the hexagon shown in Fig. \ref{fig:hexagon}, then it can be divided up by
null lines into a collection of intervals and causally convex triangles
such as Fig. \ref{fig:triangle}. Then the formula (\ref{eqn:action_euler})
will apply if the mean of the Discrete Action for a causally convex triangle
tends to $1$ in the infinite density limit.

\begin{figure}[htb]
        \centering \subfloat[causally convex triangle]{
                \includegraphics[scale=1.2]{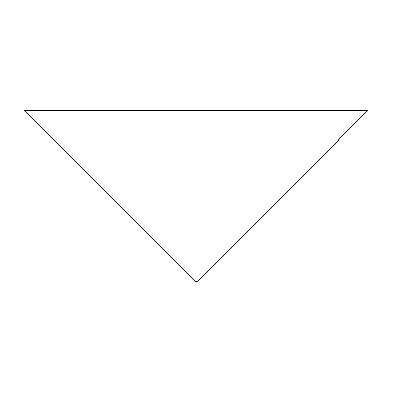}
                        \label{fig:triangle}
                }
                \hskip 1.5cm \subfloat[causally convex hexagon]{
                \includegraphics[scale=1.25]{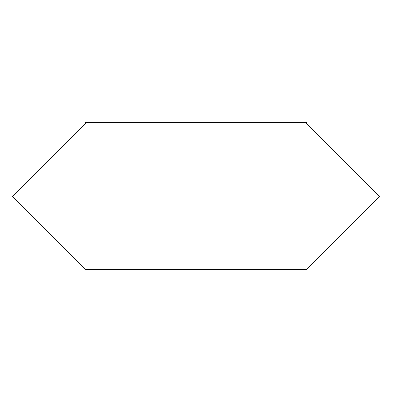}
                        \label{fig:hexagon}
                }
        \caption{Causally convex regions with boundaries formed from null and spacelike line segments}
        \label{fig:trihex}
\end{figure}

First note that by Poincar\'e invariance we can choose coordinates so that 
the spacelike edge of the triangle is at $t=$constant, and the apex
lies at the origin. 

Using null coordinates, as before, we have
\be\label{eq:action_triangle}
\langle \bS_\triangle \rangle = 
\rho V - 2\rho^2 \int_0^L dv_x \int_0^{v_x} du_x
\int_0^{v_x} dv_y \int_0^{u_x} du_y \, p(\rho \Delta u \Delta v) 
\ee
where $L = \sqrt{2 V}$ and $V$ is the area of the triangle. 
This gives
\be
\langle \bS_\triangle \rangle = 1 + \frac{1}{\rho V} + 
O\left((\rho V)^{-2}\right) \approx 1\,.
\ee
We see that the mean DA of the triangle
does indeed tend to $1$ as $\rho \rightarrow \infty$, 
though the corrections are not exponentially 
small.

Now, consider a general causally convex region with a boundary whose spacelike 
portion is curved. So long as the 
discreteness scale is small enough -- small compared to the 
radius of curvature of the boundary --  we can 
tile the region with intervals and with  causally 
convex approximate triangles along the spacelike
boundary, all of which are large enough compared to the 
discreteness scale for the Formula (\ref{eqn:action_euler}) 
to hold approximately. We conclude that the mean of the DA for 
any causally convex region of $\bbM^2$ will tend to $1$ in the 
limit of infinite density.

Is causal convexity  necessary for the
mean of the DA to be approximately 1? When a region, $R\subset \bbM^2$, is 
not causally convex, there will exist pairs of points
$x,y \in R$ such that the causal interval in $R$ between $x$ and $y$
is smaller than the causal interval 
between them in $\bbM^2$ (the ``diamond''). Since it is the volume of the
 causal interval in $R$
which appears in the expression for the mean of the DA, one might expect this
to disrupt the result and indeed it does.

Consider the Discrete Action, $\bS_\oblong$ of 
a rectangle with edges parallel to the $t$ and $x$
axes. Analytic computation of the expectation value $\langle \bS_\oblong
\rangle$ is hard exactly because of the lack of 
causal convexity: the integral (\ref{eq:meanS})
 breaks up into several subintegrals
depending on the positions of $x$ and $y$ relative to the
boundary. Therefore we use simulations to estimate the 
value. A sprinkling into a
rectangle has three independent parameters that fully characterise 
the problem. One choice is the spatial width $w$, the height
along the time-axis $h$ and the sprinkling density
$\rho$.\footnote{Width $w$, height $h$ and expected number of sprinkled
elements $N$ would be another choice.} 
The expectation value $\langle \bS_{\oblong,w,h,\rho} \rangle$
must be invariant under rescaling
\begin{equation}
\label{eqn:action_square_scaling}
\begin{array}{rcl}
w & \rightarrow & \lambda\cdot w \\ h & \rightarrow & \lambda\cdot h
\\ \rho & \rightarrow & \lambda^{-2}\cdot \rho.
\end{array}
\end{equation}
\begin{figure}[htb]
\hspace{1.4cm}  
\centering \subfloat[Simulation data for the action of a
                rectangle in $\mathbb{M}^2$ for $w=h=1$, varying
                density $\rho$ with a power-law fit. Data averaged
                over $10^6$ to $10^7$ runs. Fit function: $\rho^a
                \cdot b$.]{
                \includegraphics[scale=1.1]{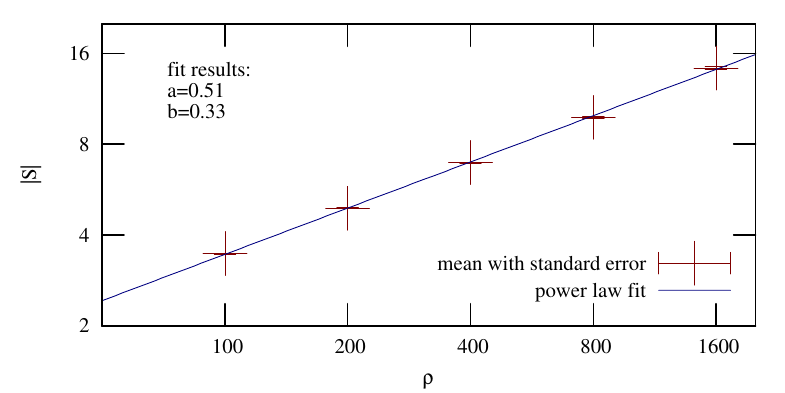}
                        \label{fig:action_square_sim1}
                }
\newline                
\subfloat[Simulation data for the action
                of a rectangle in $\mathbb{M}^2$ for $w=1, \rho=100$,
                varying height $h$ with a power-law fit. Data averaged
                over $10^6$ to $10^7$ runs. Fit function: $h^a \cdot
                b$.]{
                \includegraphics[scale=1.1]{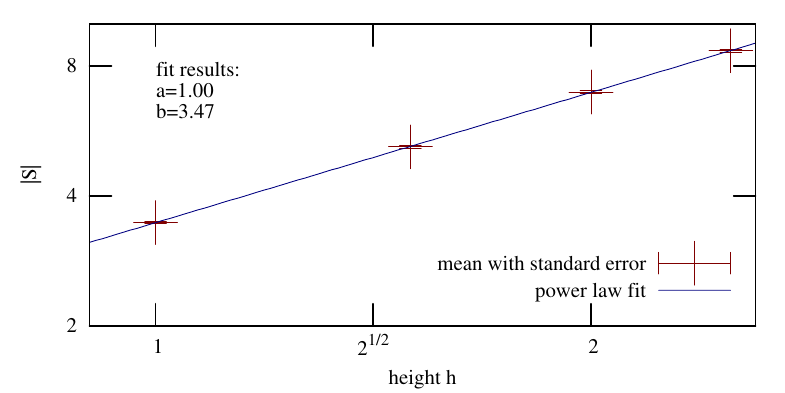}
                        \label{fig:action_square_sim2}
                }
        \caption{Numerical results for the action of a rectangle in
        $\mathbb{M}^2$.}
        \label{fig:action_square_sim}
\end{figure}
Fig. \ref{fig:action_square_sim} shows simulation data for two
different setups with power-law
fits. Fig. \ref{fig:action_square_sim1} shows $\langle \bS \rangle$ for
constant $w$ and $h$ and varying $\rho$,
Fig. \ref{fig:action_square_sim2} for constant $w$ and $\rho$ and for
varying $h$. Given the small relative error bars the power-law fits
look quite convincing and we will assume that $\langle \bS_\oblong
\rangle$ can, at least in the regime covered by the simulations, be
written in the form
\begin{equation}
\label{eqn:action_square_powerlaw}
\langle \bS_\oblong \rangle =\text{const} \cdot h^\alpha w^\beta \rho^\gamma.
\end{equation}
The  scale invariance (\ref{eqn:action_square_scaling}) demands $\alpha+\beta-2
\gamma=0$. From simulation 1 (Fig. \ref{fig:action_square_sim1}) one
is tempted to deduce $\gamma=1/2$ and from simulation 2
(Fig. \ref{fig:action_square_sim2}) that $\alpha=1$. It follows
$\beta=0$.

The fact that for constant $\rho$ the width does not affect the value
of the action whereas $\langle \bS_\oblong \rangle \propto h$
suggests that in general $\langle \bS_\oblong \rangle$ contains 
boundary terms from timelike boundaries only. We return to this
question in the discussion section.


\section{The flat cylinder}

In order to apply formula Eq. (\ref{eqn:action_euler}) to a causal
interval, $\cI_c$, of height $T$ on a cylinder with circumference $L$ with
$L\leq T \leq 2L$ one might come up with a tiling into 
subintervals, $\cI_i$, $i = 1,\dots 8$, as shown in
Fig. \ref{fig:action_cylinder_split}.
\vspace{0.3cm}
\begin{figure}[htb]
        \centering {
                        \includegraphics[scale=1.2]{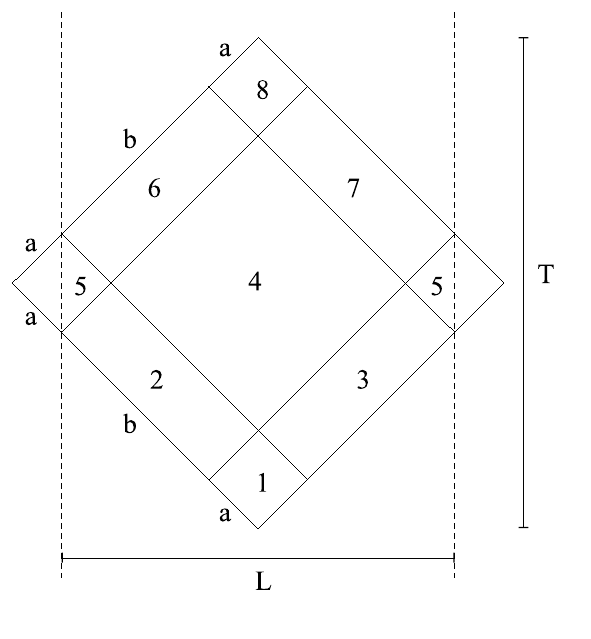}
                }
                      \caption{Tiling of the interval $\cI_c$ with $L\le T\le 2L$.
                      $a$ and $b$ are the $u$ and $v$ coordinate
                      lengths of the sides of the subintervals shown.}
                       \label{fig:action_cylinder_split}
    \end{figure}
Taking into account the topological identification, we have
$F=8, E=12, V=4$ thus yielding a predicted
high-density expectation value of $\langle \bS_{\cI_c} \rangle \approx 0$.
 However
we have not shown yet that formula Eq. (\ref{eqn:action_euler}) is
applicable to the cylinder. The division of the causal
interval in Fig. \ref{fig:action_cylinder_split} has been chosen such
that formula Eq. (\ref{eqn:action_minkowski2}) for the faces and
formulae Eq. (\ref{eqn:action_minkowski_21}) and
(\ref{eqn:action_minkowski_41}) for the bilocal contributions of two
intervals that share an edge or lie above and below a 
vertex can still be applied as the
cylinder topology does not affect these cases. But the computation
of contributions like (5,1),(6,1) and (8,1) differs from the Minkowski
setup due to the nontrivial topology.

Recall
\be
\langle \bS_{\cI_c} \rangle = \rho V -2K
\ee
where
\be 
K = \rho^2 \int_{\cI_c} d^2y \int_{\cI_c \cap J^+(y)} d^2x \,\,p(\rho V_{xy})\,.
\ee
In general, 
$K$ can be split into a sum of terms, $K = \sum_{\alpha=1}^{\infty} K_\alpha$ depending on how many homotopy classes of causal curves there are from $y$ to $x$:
 \be
 K_\alpha := \rho^2 \int_{\cI_c} d^2y \int_{\cI_c \cap J_\alpha^+(y)} d^2x \,\,p(\rho V_{xy})
 \ee
 where
 \be 
 J_\alpha^+(y) := \{x \in J^+(y)\,|\, \exists \ {\textrm{ exactly $\alpha$ homotopy classes of
 causal curves from $y$ to $x$}}\}\,.
 \ee
This split is motivated by the fact that $V_{xy}$ strongly depends on the number of homotopy classes of causal paths between $x$ and $y$. For our interval, $K_\alpha = 0$ for $\alpha > 3$. 

From Fig. \ref{fig:action_cylinder_split} we see the relation between $a,b,T$ and $L$ is:
\begin{eqnarray}
a & = & (T-L)/\sqrt{2} \nonumber \\ b & = & (2L-T)/\sqrt{2}
\end{eqnarray}
The causal volume $V_{xy}$ for $x\in J_\alpha^+(y)$ for $\alpha \ge 3$ is at least $(a+b)^2$ so 
$K_3$ is suppressed by at least $\exp(-\rho(a+b)^2)$ and can thus be neglected as $L = \sqrt{2}(a+b)$ is assumed to be large in discreteness units of $\rho^{-\frac{1}{2}}$.

The values for $K_1$ and $K_2$ are \cite{Schmitzer:2010}
\bea
K_1 & = & \frac{\rho V}{2}+\frac{1}{2}\exp(-\rho a^2)-(1+\rho a b)\exp(-\rho a(a+b)) + \text{corr.} \nonumber \\
K_2 & = & - \frac{2}{(a+b)^2\rho}+\exp(-\rho a (a+b))\left[1+\rho a b\right.\nonumber \\
& & \left. +\frac{1}{(a+b)^4 \rho^2}\left((6+2\rho(a+b)(2a+b)-\rho^2 (a+b)^2 b^2+\rho^3 (a+b)^3 a b^2) \right. \right. \nonumber \\
& & \left. \left.-2 \exp(-\rho a (a+b))(3+4 \rho a (a+b)+2\rho^2 a^2 (a+b)^2) \right) \right]+ \text{corr.}
\eea
where ``$+$ corr.'' stands for neglected terms suppressed by $\exp(-\rho(a+b)^2)$. However we 
will keep terms with factors  $\exp(-\rho\,a^2)$ and $\exp(-\rho\,a\,(a+b))$ since for $T$ only
slightly larger than $L$ the value of $a$ will be very small and these terms are then significant. 

The overall action is
\bea
\label{eqn:action_cylinder_result}
\langle \bS_{\cI_c} \rangle & = & -\exp(-\rho a^2)+2(1+\rho a b)\exp(-\rho a(a+b))\nonumber \\
& & + \frac{4}{(a+b)^2\rho}+\exp(-\rho a (a+b))\left[1+\rho a b\right.\nonumber \\
& & \left. +\frac{1}{(a+b)^4 \rho^2}\left((6+2\rho(a+b)(2a+b)-\rho^2 (a+b)^2 b^2+\rho^3 (a+b)^3 a b^2) \right. \right. \nonumber \\
& & \left. \left.-2 \exp(-\rho a (a+b))(3+4 \rho a (a+b)+2\rho^2 a^2 (a+b)^2) \right) \right]+ \text{corr.}\ .
\eea

For $T>2L$ consider a division of the interval into regions 1 and 2 as shown in Fig. \ref{fig:action_cylinder_long}. The expected DA is the sum of the expected actions for regions 1 and 2 and the bilocal contribution $\langle \bS_{21}\rangle$.

\begin{figure}[htb]
	\centering
		{
			\includegraphics[scale=1.3]{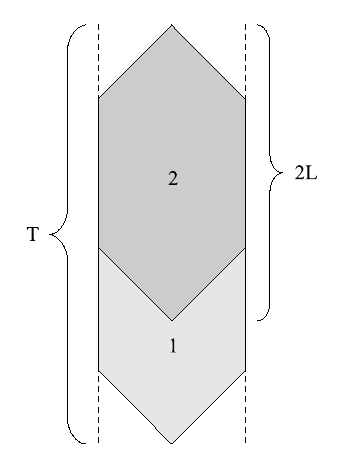}
		}
	\caption{Division of interval when $T>2L$.}
	\label{fig:action_cylinder_long}
\end{figure}
It can be shown \cite{Schmitzer:2010} that the expected action 
for region 1 and the bilocal contribution cancel (up
to exponentially small terms) and the result is 
just given by the expected action of region 2 which can be obtained from Eq.(\ref{eqn:action_cylinder_result}) by setting $a=L/\sqrt{2},b=0$ (and now neglecting all exponentials as $a$ is no longer close to 0):
\be
\langle \bS_{\cI_c}\rangle  = 
\frac{8}{L^2 \rho} + {\text{corr.}} \,.
\ee
Fig. \ref{fig:action_cylinder_simulation} shows a plot of the
 analytic expectation value for the cylinder action compared to simulation results.
For $T\rightarrow L$ the action approaches the
Minkowskian limit $1$. For $T$ only slightly greater than $L$ the
exponential terms
dominate and cause a downwards spike. As $T\rightarrow 2L$ the non-exponential
correction, $\frac{8}{\rho L^2}$ (which comes from $K_2$) dominates.
However this also tends to zero in the limit $\rho\rightarrow \infty$ so 
 $\langle \bS_{\cI_c}\rangle \approx 0$ as initially predicted.
Indeed it can be shown explicitly that the bilocal contributions from pairs
of intervals that do not share and edge or vertex tend to zero as $\rho\rightarrow \infty$
and so the formula $F - E + V$ can be applied to intervals of the cylinder. More generally, 
the previous argument regarding null tilings of causally convex regions of $\bbM^2$ can be given
here, and we conclude that 
$\langle \bS\rangle \approx 0$ for general topologically non-trivial
causally convex regions of the cylinder. 

\begin{figure}[htb]
        \centering {
                \includegraphics[scale=1.1]{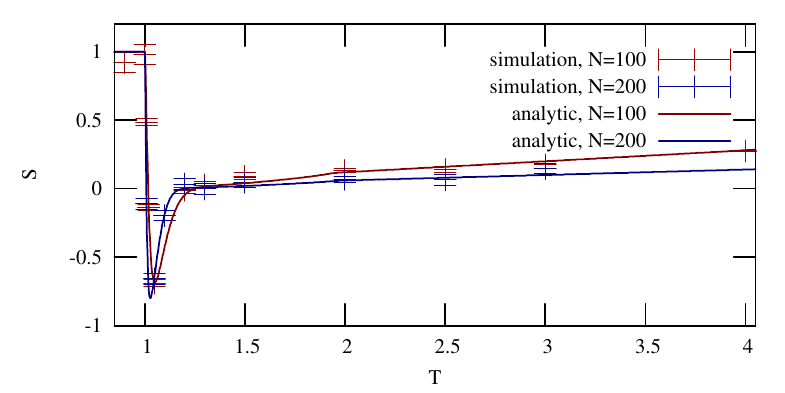}
                }
        \caption{The expected action of a cylinder-interval for $L=1, \langle N \rangle=100$ and $\langle N \rangle = 200$ compared with simulation results.}
        \label{fig:action_cylinder_simulation}
\end{figure}

\section{The flat trousers}

We investigate now a causally convex neighbourhood of the 
flat 1+1 trousers spacetime in which two $S^1$'s join
to form a single $S^1$. The trousers spacetime is a piece of $\bbM^2$ with 
	cuts and identifications as shown in 
Fig. \ref{fig:trousers}. Although the singularity, $P$, at 
which the topology changes is by some definitions not 
strictly in the spacetime since the metric degenerates there, 
nevertheless the causal order is well defined at 
the singularity: it is 
clear what the causal past and causal future of $P$ are.
Therefore we will consider $P$ as a point of the manifold. 
Note that in any sprinkling into the trousers almost 
surely no element will be sprinkled at $P$. 

\begin{figure}[htb]
	\centering
		{
			\includegraphics[scale=1.3]{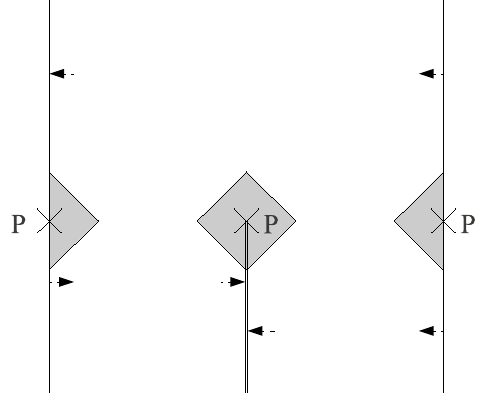}
		}
	\caption{The trousers spacetime. 
	$P$ is the singularity -- all three instances of $P$ are identified --
	and the shaded region is a neighbourhood 
	of $P$. There is a vertical cut down from the central 
copy of $P$ with the two legs identified as shown.}
	\label{fig:trousers}
\end{figure}

Let  $\cN$ denote the neighbourhood of $P$ shown
as the shaded region in 
Fig.  \ref{fig:trousers}. It  consists of 
two flat intervals each with $P$ as their
midpoint, identified across ``branch cuts'' from $P$ to their
past tips.  $\cN$  is topologically a disc  if $P$
is in included the manifold and if the formula (\ref{eqn:action_euler}) 
holds then the expected DA of $\cN$ would be 
equal to 1 in the limit of large density. 

Let the volume (area) of each of the two 
intervals be $4a^2$ and consider the null tiling into 
8 intervals, $\cI_i$, $i = 1,\dots, 8$, 
shown in  Fig. \ref{fig:crotch1}. The interval $\cI_1$ 
comprises the two triangles labelled $1'$ and $1''$ and the 
interval $\cI_2$ comprises the triangles labelled $2'$ and $2''$.
Adopting the same notation for the bilocal discrete action of 
two intervals used in (\ref{eqn:action_minkowski_split}) we have
\be
\langle \bS_\cN \rangle 
= \sum_{i=1}^8 \langle \bS_{\cI_i}\rangle 
+  \sum_{\substack{i,j =1\\ j<i}}^8 \langle
\bS_{ij}\rangle\,.
\ee

\begin{figure}[htb]
	\centering
		{
			\includegraphics[scale=1.3]{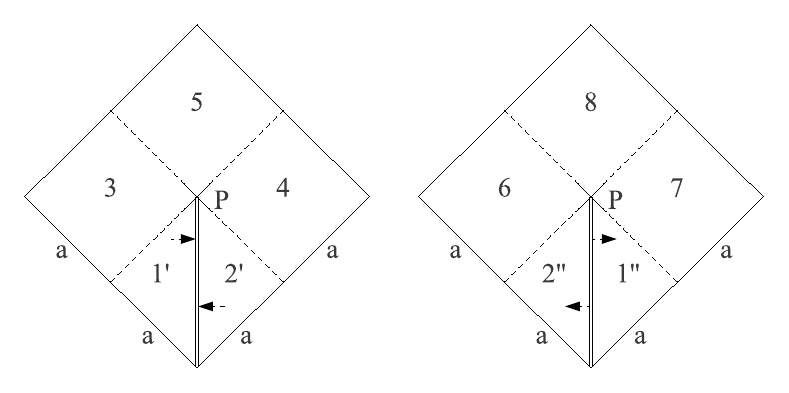}
		}
	\caption{Null tiling of $\cN$ into 8 intervals.}
	\label{fig:crotch1}
\end{figure}

For each $i$, $\langle \bS_{\cI_i}\rangle \approx 1$.
The bilocal terms are nonzero when the intervals
$\cI_i$ and $\cI_j$ share an  edge and in that 
case $\langle \bS_{ij}\rangle \approx -1$. There are 
8 edges so these contributions cancel the 
contributions of the 8 individual intervals. The only 
other nonzero bilocal terms are 
$\langle \bS_{ij}\rangle$ where $i = 5,8$ and $j = 1,2$
and their sum is the contribution of the vertex at the 
singularity. 
These  4 terms are equal by symmetry so we have
$\langle \bS_{\cN}\rangle = 4\langle \bS_{51}\rangle $.

The causal interval between $x\in \cI_5$ and
$y\in \cI_1$ is shown in Fig. \ref{fig:crotch2} 
and we deduce that 
\be
\langle \bS_{51}\rangle = 
-2  \int_a^{2a} du_x \int_a^{2a} dv_x
 \int_{0}^{a} du_y \int_{0}^{a} dv_y\,\rho^2\,p(\rho V_{xy})\,. \label{eqn:action_crotch} 
\ee
where
\be
V_{xy} = \Delta u \Delta v - (v_x -a)(a - u_y)\,.
\ee

\begin{figure}[htb]
	\centering
		{
			\includegraphics[scale=1.3]{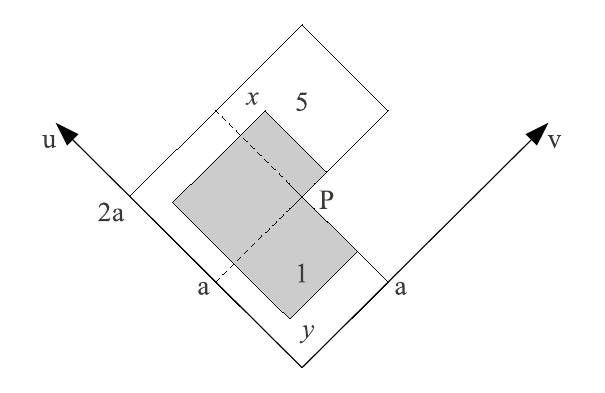}
		}
	\caption{The causal interval between $x\in \cI_5$ and
$y\in \cI_1$ is depicted in grey.}
	\label{fig:crotch2}
\end{figure}

This gives
\be \label{eqn:action_crotch_result}
\langle \bS_{\cN}\rangle = 4 \ln(\rho a^2) + 4(\gamma -1) + {\textrm{O}}\left(\frac{1}{\rho a^2}\right) 
\ee
where $\gamma$ is Euler's constant.
We see that the expected DA of the neighbourhood of the 
singularity does not tend to $1$ or any constant but
grows logarithmically with the density.



\section{Discussion}

We have shown that in the limit of infinite
density, the mean of the Discrete Action will be 1 for any causally
convex region of $\mathbb{M}^2$ including regions whose past and/or
future boundaries contain spacelike segments. Since these spacelike
segments may have nonzero geodesic curvature, the constancy of the
mean of the DA suggests that it contains no contribution from the
past or future boundaries.

Indeed a handwaving argument can be given as to why this should be so, 
even when $\cM$ is curved. The boundary of a causally convex region
$\mathcal{U} \subset \cM$ 
consists of a future boundary and a past boundary intersecting 
in a co-dimension 2 spacelike surface. There is no timelike 
portion of the boundary. The mean of the DA 
 is a double integral over $\mathcal{U}$ which can be done in
either order. The integrand is a retarded 2-point function,
\be 
\rho {\mathcal{L}}(x,y) = \frac{\rho}{\sqrt{-g}} \delta^{(2)}(x,y) - 2 \rho^2 \,p(\rho V_{xy}) C(x,y)
\ee
where $C(x,y) = 1$ if $y \in J^-(x)$ and $0$ otherwise. 
Let us assume the density is high enough that a sprinkled causal
set can capture the curvature of $\cM$, {\textit{i.e.}} at 
each point $y \in \cM$ there is a local inertial frame in which 
the curvature components are small compared to the density. 
 If we do the $x$ integration first,
at fixed $y$, then the resulting function $\rho L(y)$ is
approximately $\frac{1}{4}R(y)$ unless $y$ is too close to the future
boundary. If it is within length $\rho^{-\frac{1}{2}}$
of the boundary then the range of the $x$  
integration will  not be large enough for the
approximation to hold \cite{Benincasa:2010ac}. 
Then we do the integration over $y$ to get approximately the
usual Einstein-Hilbert bulk term together possibly with some contribution from the
integral over the points $y$ close to the future boundary, {\textit{i.e.}} possibly
some kind of future boundary term. But there is no contribution from
the past boundary at all.  Now reverse the order of integration: do $y$
first and then $x$. Now there appears to be no contribution from the
future boundary.  This can only happen if neither boundary
contributes. So the only points where some boundary contribution can
come in, is from the points which are close to both past and future
boundaries {\textit{i.e.}}  from the spacelike co-dimension 2 ``corners'' where the
past and future boundaries intersect.  The argument holds when the
past and future boundaries are partly spacelike as well as when they
are wholly null. There is no reason, from this argument, that timelike 
boundaries could not contribute however and we saw evidence that they 
do from the results for the rectangle. 
 
 This heuristic reasoning would have to be backed up with further evidence from simulations
 of the Discrete Action
 but it suggests that there is no Gauss-Bonnet formula for the 2D Discrete Action. 
 The 2D Gauss-Bonnet Theorem can hold
 because, as the geometry of the bulk surface is varied, the extrinsic curvature of the boundary
changes and the right combination of bulk and boundary terms can remain constant. 
In 2D the co-dimension 2 ``corner'' is  an $S^0$, {\textit{i.e.}} 2 points, and
if the only boundary contributions are from these 2 points,
these couldn't compensate for the changing bulk 
term. Another reason not to expect the DA to 
satisfy a Gauss-Bonnet formula is that it appears that the 
appropriate Lorentzian analogue of the Euclidean formula 
is of the form ``bulk term + boundary term + corner terms'' $=2 \pi i \chi$ rather than $2\pi \chi$
\cite{Law:1992,Louko:1997jw} (see also \cite{Birman:1992,Jee:1992}).
Both the bulk and boundary terms are real but the formula can hold  
because the corner contributions are Lorentzian angles which can be 
complex. However, the Discrete Action is real.

This putative lack of boundary terms 
could explain why the expected DA for any causally convex 
region of $\bbM^2$ is the same. The continuum 
bulk term is zero. If the mean of the DA is indeed close to 
the continuum bulk term plus only a contribution from the $S^0$ corners
then that should be the same for all causally convex regions. 
Presumably, the difference for the neighbourhood of the singularity of the 
trousers comes from a boundary effect of the non-standard causal 
structure around the singularity which has a double lobed past and future. 
These issues all remain to be investigated. 

There are a large number of open questions. 
What does happen in 2D curved spacetimes? Will the results bear out the conjecture that the 
expected DA is approximately the Einstein-Hilbert term plus a constant from the $S^0$
corner? What happens in higher dimensions? There are analogues of the 2D Discrete Action
in 4D \cite{Benincasa:2010ac} and 3,5,6,7D and higher \cite{Glaser:2010}.
One would expect, for example, that the mean of the DA of an interval in $\bbM^d$  
would be proportional to the volume of the $S^{d-2}$ corner.

 Although
 one need not take any position on quantum gravity 
 to find interest in the Discrete Action as a random variable defined
for a continuum spacetime -- one need not consider the discreteness of the 
causal sets that arise in the definition of the Discrete Action to have any 
physical basis -- its main application is likely to be in the causal set 
approach to quantum gravity. So, what is the 
significance in quantum gravity of the results for the interval, trousers and 
rectangle? For example, the result for the rectangle suggests that the expected
DA contains boundary contributions
proportional to the length of any timelike boundary.  Can we use the 
DA to give an argument against 
the appearance of ``holes'' and ``edges'' in spacetime? Or 
for or against topology changing processes such as the trousers?

A major open question is how the fluctuations in the DA behave as the
density gets large: we should stress that the results reported here are all
concerning the mean of the DA. For a typical sprinkled causet, how far is the 
DA from the mean? Preliminary results show the fluctuations grow as the 
density gets large \cite{Schmitzer:2010}, contrary to the hope 
expressed in \cite{Benincasa:2010ac} and this needs to be 
studied further. To tame the fluctations it may be necessary to introduce 
a mesoscale between the 
discreteness scale and the observation scale \cite{Sorkin:2007qi, Benincasa:2010ac}.
Further work is needed to illuminate these issues. 

\section{Acknowledgments}
We are grateful to Graham Brightwell, David Rideout, Rafael D. Sorkin and Sumati Surya for
stimulating and useful discussions. FD and DMTB are partially supported by a Royal Society Grant IJP 2006/R2
and thank Raman Research Institute, Bangalore, India and 
 the Perimeter Institute for Theoretical Physics, Waterloo, Canada for hospitality 
whilst carrying out this work.  

\bibliography{refs}  \bibliographystyle{JHEP}

\end{document}